\documentclass[twocolumn,aps,pre]{revtex4}
\usepackage{graphicx}

\newcommand{\Eqn}[1]{Eq.~(\ref{#1})}     
\newcommand{\Sec}[1]{Section~\ref{#1}}     
\newcommand{\Fig}[1]{Fig.~\ref{#1}}     
\newcommand{\Figs}[1]{Figs.~\ref{#1}}     
\newcommand{\LN}[1]{\ln \left( #1 \right)}
\newcommand{\aver}[1]{\left< #1 \right>}

\newcommand{\zed}{z}
\newcommand{\lden}{d_{l}}
\begin{document}

\title{Yard-Sale exchange on networks: Wealth sharing and wealth
  appropriation.}

\author{R.~Bustos-Guajardo and Cristian F.~Moukarzel}

\affiliation{CINVESTAV del IPN, Depto.~de F\'{\i}sica Aplicada\\97310 M\'erida, Yucat\'an, M\'exico.}

\begin{abstract}
  Yard-Sale (YS) is a stochastic multiplicative wealth-exchange model
  with two phases: a stable one where wealth is shared, and an
  unstable one where wealth condenses onto one agent. YS is here
  studied numerically on 1d rings, 2d square lattices, and random
  graphs with variable average coordination, comparing its properties
  with those in mean field (MF). Equilibrium properties in the stable
  phase are almost unaffected by the introduction of a
  network. Measurement of decorrelation times in the stable phase
  allow us to determine the critical interface with very good
  precision, and it turns out to be the same, for all networks
  analyzed, as the one that can be analytically derived in MF. In the
  unstable phase, on the other hand, dynamical as well as asymptotic
  properties are strongly network-dependent. Wealth no longer
  condenses on a single agent, as in MF, but onto an extensive set of
  agents, the properties of which depend on the network. Connections
  with previous studies of coalescence of immobile reactants are
  discussed, and their analytic predictions are successfully compared
  with our numerical results.
\end{abstract}
\maketitle
\section{Introduction}
\label{sec:intro}
Wealth exchange models, initially proposed to investigate the
emergence of wealth inequality~\cite{PCDP96} in human societies, have
recently became a subject of intense research~\cite{CYCEWD05,YRCSM09},
following the availability of massive amounts of statistical data
describing commercial exchange, as well as wealth and income
distributions in different contexts~\cite{Note1}.
\\
Conservative stochastic exchange models were first used by
Angle~\cite{ATST86}, who considered the spontaneous buildup of wealth
differences among equally able agents. In Angle's initial model,
wealth concentration is a consequence of an explicit statistical
advantage favoring richer agents. In other words, the ``rich get
richer'' phenomenon is assumed explicitely in the exchange rules.
Later work showed~\cite{HFTM02,SSMW03,IGACBR04,MGIWCI07,MMAE11} that
an explicit advantage favoring the rich is not necessary for wealth
concentration to appear.  Wealth concentration can develop even if the
poor have an explicit statistical advantage. This rather
counterintuitive result, which has only recently been
stressed~\cite{MGIWCI07,MMAE11} in the Econophysics literature, arises
when the amount at stake in each transaction is proportional to the
poorest agent's wealth, e.g~in the so-called Yard-Sale (YS)
models~\cite{IKRWDI98,HFTM02,SSMW03,IGACBR04,SPWATM04,MGIWCI07,MMAE11}.
Yard Sale is an example of Multiplicative Stochastic Exchange, so
named because the wealth of the poorest intervening agent is
multiplied by a random number after the exchange~\cite{MMAE11}.  Under
YS dynamics, in the long run all wealth may end up in the hands of one
lucky agent, even if each pairwise transaction is statistically biased
in favor of the poorest of the two intervening agents. Therefore,
favoring the poor may not suffice to avoid wealth concentration, if the
bias in their favor is not strong enough. Interestingly, YS rules
constitute a realistic (although highly simplified) microscopic model
for the wealth exchange process occurring during commercial
interaction, or trade~\cite{HFTM02,SSMW03,SPWATM04}. This suggests
that the conditions for the spontaneous creation of enormous wealth
differences for no reason other than luck~\cite{TF01}, are built into
the commercial exchange rules, even if these rules may superficially
appear to favor the poor. Because of the possibility of
counterintuitive properties such as this one, and because of their
relevance for real world commercial exchange, it is clearly of
interest to understand the phenomenology of multiplicative exchange
models thoroughly.

In simple versions of YS, pairs of agents 'bet' for a fraction
\hbox{$f\leq 1$} of the wealth of the poorest of them, who has a
probability $p$ to win the bet.  Depending on $p$ and $f$, long-term
evolution can give rise either to a nontrivial equilibrium wealth
distribution $P(w)$ or to \emph{condensation} of the whole wealth in
the hands of just one agent.  We call the resulting phases,
respectively the wealth-sharing (or stable) and the
wealth-appropriation (or unstable) phase.  To date, most results for
this model concern the full-mixture (or Mean-Field) case. However,
commercial exchange is often determined by geographical, social or
other constraints, which are ignored in the fully mixed
approximation. Usually, a given agent can only exchange wealth with a
reduced subset of other agents who are ``close'' to him by some
measure of distance. These constraints can be described, at the
simplest level, by means of a network in which nodes $i=1,2,\ldots,N$
are economic agents and edges $ij$ represent their possible
interactions. It is reasonable to expect the topological properties of
this network of allowed interactions to have a strong impact upon the
general properties of wealth exchange processes occurring on them.

Recent work~\cite{SBTOT03,GLFTP04,GD-MAIBT07} explores the network of
commercial interactions among nations, or ``World Trade
Web''~\cite{SBTOT03}. These studies make it clear that the topology of
interaction networks is strongly correlated with the dynamical and
static properties of the resulting wealth exchange process taking
place on those networks~\cite{Note2}.  Numerical
investigations~\cite{SFAS01,GLWDO04,GLEON08} of the
Bouchaud-Mezard~\cite{BMWCI00,SPWATM04} model (BMM) on networks, find
wealth distributions $P(w)$ that change from lognormal to power-law
when the connectivity is increased, for reasons that are easily
understood. However, the BMM, while being an interesting solvable
model, considers linear exchange (the amount exchanged is proportional
to wealth differences), which is not very realistic~\cite{SPWATM04}.
On the other hand, the BMM includes both exchange and nonconservative
processes (such as investment), and it is precisely from the interplay
between the two that the network's connectivity becomes important in
the above studies.

The aim of this work is to analyze network effects for a realistic
model of pure conservative commercial exchange.  Although real
economic systems involve nonconservative wealth-modifying processes as
well, it is important to first understand the properties of the
individual wealth-affecting mechanisms in isolation. In this work, we
study the Yard-Sale (YS) model on networks. Our network-restricted
version of YS is defined as follows: At each timestep, every agent $i$
interacts (exchanges wealth according to YS rules) with another agent
$j$ that is randomly chosen among its neighbors, i.e. among those
agents $j$ for which a link $ij$ exists.  The interaction network is
fixed in time. Therefore, a pair of agents not connected by a link
will never interact directly.  If the coordination number $\gamma$ is
roughly the same for all nodes, each agent engages in two interactions
per timestep, on average. We consider here one-dimensional chains and
two-dimensional square lattices with nearest-neighbor links and
periodic boundary conditions, as well as Erd\H{o}s-R\'enyi Random
Graphs~\cite{BRG01} with variable coordination $\gamma$. In the limit
$\gamma \to (N-1)$, the Random Graph becomes a complete graph, and
every pair $ij$ has the same probability to interact. This is the
full-mixture case.
\\
We focus on the identification of network-specific effects, i.e.~the
extent to which the static and dynamic properties of the YS model,
when implemented on a network, depart from those in full-mixture.  Our
results show that, while the stable wealth distribution $P(w)$ is
mildly network-dependent, the location of the interface $p^{*}(f)$
that delimits the stable phase remains the same as in the full-mixture
case, for all networks considered in this work.  The critical line
$p^{*}(f)$ is therefore universal, in the sense defined in the context
of the theory of phase transitions.  Dynamical properties, as for
example decorrelation times, on the other hand, do depend on the
network. Decorrelation times, which in this case are a measure of
``social mobility'' of agents, are found to diverge at the interface
with the unstable phase. This divergence is used to locate the
critical line with high precision.

In the unstable, or wealth-appropriation, phase, dynamical as well as
long-time properties of YS are found to be strongly network-dependent.
The most important difference with full-mixture is in this case that,
on a network, complete condensation in the hands of one agent no
longer occurs. On a network, instead, in the long run an extensive set
of ``locally rich'' agents (LRA) appears, each connected only to
extremely impoverished agents. This leads to the effective cessation
of all exchange activity, a phenomenon similar to dynamical freezing.
This freezing onto a disordered final state is observed in the whole
unstable phase. The properties of the final set of LRAs, and their
final wealth distribution, depend on the network topology, as well as
upon $p$ and $f$. We discuss the connections between the appearance of
a set of locally rich agents and the process of coalescence (or
coagulation) of immobile reactants~\cite{BVNR99,AGNN01,AO04} on
networks. These connections provide analytical predictions for the
number of LRAs on Random Graphs, which are consistent with our own
numerical results. By increasing the average connectivity $\gamma$ of
a network, the number of LRAs onto which wealth condenses is
decreased, until, in the limit $\gamma = N-1$, which is the
full-mixture case, only one LRA remains, i.e~full condensation is
recovered.

This article is organized as follows. \Sec{sec:yard-sale-full} recalls
some results for YS in full-mixture. In
\Sec{sec:numer-results-stable}, numerical results on networks in the
stable phase are presented and compared with full mixture results. In
particular, decorrelation times are used in this section to locate the
interface with high precision.  Wealth appropriation dynamics on
networks is studied in \Sec{sec:numer-results-unst}, where it is found
that wealth condenses onto an extensive number of locally rich agents
(LRA). Their number and wealth distribution are analyzed in this
section. Finally, \Sec{sec:discussion-results} offers a discussion of
our results.
\section{Yard-Sale in full-mixture}
\label{sec:yard-sale-full}
Consider wealth exchange for a pair of agents $i$ and $j$ with
$w_i<w_j$ before interaction, according to the following YS rules. The
agents bet for an amount \hbox{$f \times MIN(w_i,w_j) =f w_i$}. The
poorest agent ($i$) wins the bet with probability $p$, in which case
$w_i \to w_i (1+f)$ and $w_j \to w_j -f w_i$, or looses the bet with
probability $(1-p)$, in which case $w_i \to w_i (1-f)$ and $w_j \to w_j
+f w_i$. The wealth of the poor agent is therefore multiplied by a
random factor $\eta$, which equals $(1+f)$ with probability $p$ and
$(1-f)$ with probability $(1-p)$. Long-term evolution under these
rules gives rise either to a stable wealth distribution $P(w)$ or to
condensation, depending on $p$ and $f$.

The location of the critical line $p^{*}(f)$ below which condensation
occurs can be derived as follows~\cite{MGIWCI07,MMAE11}. The wealth of
a very poor agent undergoes a Random Multiplicative
Process~\cite{RRMP90} with multiplier $\eta$ at each timestep.  After
a large number $t$ of timesteps, the appropriate central tendency
estimator for $w$ is its geometric average \textbf{$e^{\aver{\ln w_t}}
  = w_0 \ e^{- t \theta}$}, where
\begin{equation}
  \label{eq:2}
  -\theta=
  \aver{\LN{ \eta } } = p \ln(1+f) + (1- p) \ln(1-f).
\end{equation}
If \hbox{$\theta > 0$}, there will be a systematic transference of
wealth from poorer to richer agents. This is the wealth-appropriation,
or unstable, phase. In this phase, wealth differences among agents are
amplified in time, until the whole wealth ends up in the hands of a
single (in full-mixture) agent in the long run~\cite{MGIWCI07,MMAE11}.

If $\theta<0$, the system is in the wealth-sharing, or stable,
phase. Wealth is transferred from richer to poorer agents, which tends
to ``iron out'' wealth fluctuations. In the long run, the distribution
of wealth reaches a nontrivial equilibrium form $P(w)$, which depends
on $p$ and $f$.

By the heuristic argument above, the critical interface separating
stable and unstable phases is given by \hbox{$\theta=0$}, or
\begin{eqnarray} 
  p^{*}(f)  &=& \frac{\log(1-f)}{\log(1-f)-\log(1+f)}.
  \label{eq:17} 
\end{eqnarray}
A more rigorous analysis~\cite{MGIWCI07,MMAE11}, involving the master
equation for $P(w)$ in the full-mixture approximation, confirms
(\ref{eq:17}).
\\
Notice that the average return of the poorest agent~\cite{MMAE11} is
positive whenever $p>1/2$. There is thus a region $1/2 < p < p^{*}(f)$
where complete wealth concentration occurs, i.e.~poor agents
impoverish further, despite the average return of poor agents being
positive.
\subsection{The stable phase}
\label{sec:stable-phase}
\subsubsection{Time correlations}
\label{sec:time-correlations}
A dynamical characterization that is useful in the stable phase is the
relaxation timescale for equilibrium fluctuations.  The excess wealth
\hbox{$\Delta w_i(t)=w_i(t) - \bar{w}$}, where $\bar{w}$ is the
average per agent wealth, gives the amount by which the wealth $w_i$
of an agent $i$ departs from average at time $t$.  The correlation
function $C(\tau)$ at time $\tau$, averaged over $T$ timesteps, is
then defined as
\begin{eqnarray}
  C(\tau)=\frac{1}{N T} \sum_{t=1}^{T} \sum_{i=1}^{N}  
\Delta w_i(t) \Delta w_{i}(t+\tau).
\end{eqnarray}
We use here the normalized correlation function
\hbox{$c(\tau)=C(\tau)/C(0)$}, which equals one for $\tau=0$ and
decays as
\begin{eqnarray}
  \label{eq:12}
c(\tau) \sim c_0 e^{-\tau/\tau_0} 
\end{eqnarray}
for large $\tau$. A small value of $c(\tau)$ means that being richer
or poorer than average at a given time $t$ has little predictive power
$\tau$ timesteps later. Therefore, $c(\tau)$ measures the
``mobility'', in the wealth scale, of a typical agent over a time
horizon of $\tau$ timesteps. The timescale $\tau_0$ over which
$c(\tau)$ converges to zero measures the amount of time needed for
full ``social mixture'' (decorrelation from initial wealths, or loss
of memory). In a statistical mechanics context, $\tau_0$ is the
relaxation time needed for the decay of equilibrium
wealth-fluctuations, or ``decorrelation time''. Borrowing from the
theory of equilibrium phase transitions~\cite{SITP87,BDFTTO02}, one
expects $\tau_0(p,f)$ to diverge as the critical interface is
approached from above, as
\begin{eqnarray}
  \tau_0 (p,f) \sim {(p-p^{*}(f))}^{-\zed},
\label{eq:tau0pf}
\end{eqnarray}
where the dynamical exponent $\zed$ can be a function of $f$
eventually.  As shown later, the numerical estimation of $\tau_0(p,f)$
allows a very precise determination of the location $p^*(f)$ of the
interface.
\subsection{The unstable phase}
\label{sec:unstable-phase}
\subsubsection{Ranked wealths}
\label{sec:rank-wealths-unst}
When $p<p^{*}(f)$, the system is in the unstable phase, wealth
differences are amplified in time and this eventually leads to
condensation in the fully mixed case.  In the whole unstable phase,
the decorrelation time $\tau_0$ is infinite. Therefore an agents'
position in the wealth scale becomes frozen, in the long run. In other
words, social mobility is suppressed in the appropriation phase.

\begin{figure}[htb]
\includegraphics[width=0.70\linewidth,angle=270]{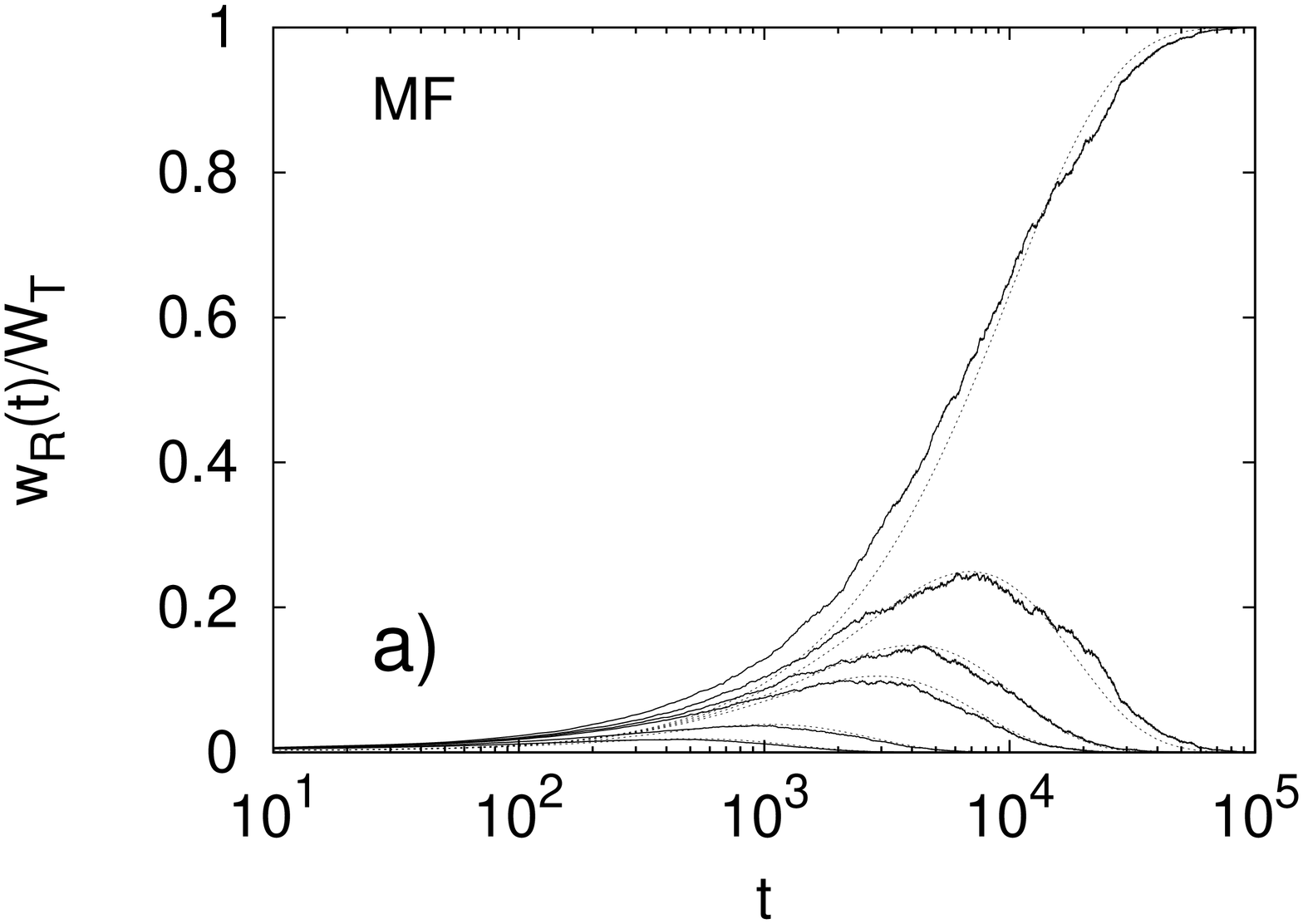}
\includegraphics[width=0.70\linewidth,angle=270]{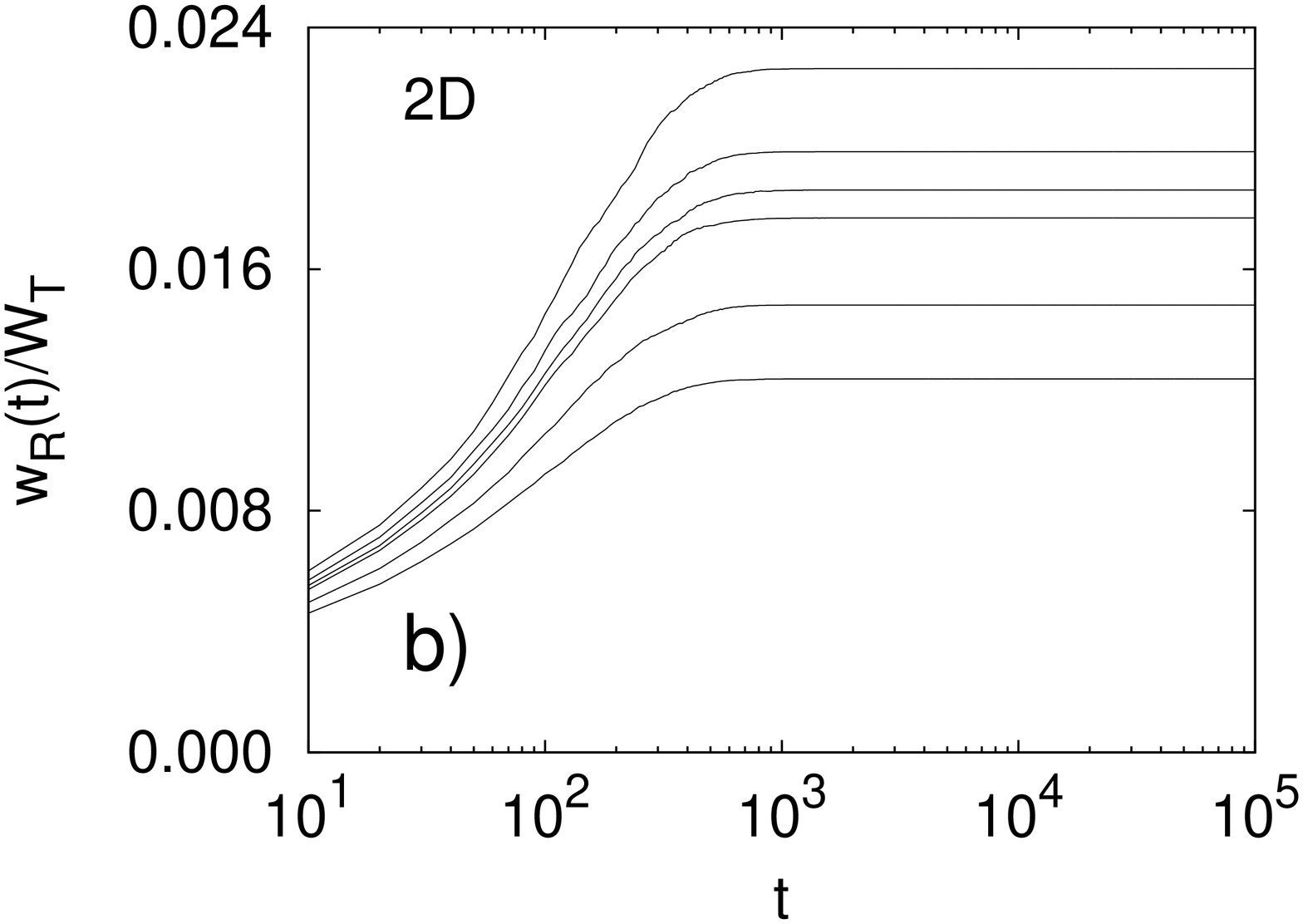}
\caption{ Time evolution, in the unstable phase, of the relative
  wealth possessed by the agent whose rank is $R$, for $R=1,2,3,4,10$
  and $20$ (from top to bottom -- full lines) from simulations in: a)
  full-mixture and, b) on a 2D square lattice with periodic
  boundaries. The dotted lines in a) show the theoretical prediction
  (\ref{eq:10}) for full-mixture. In all cases, $N=400$ agents,
  $f=0.1$, and $p=0.425$.}
 \label{fig:ginill}
\end{figure}
A dynamical analysis of the unstable phase~\cite{MMAE11} for the
fully-mixed system shows that the typical wealth $w_R(t)$ of an agent
with rank $R$~\cite{Note3} at time $t$ is \hbox{$w_R(t) \sim e^{ - t
    \theta \ \frac{(R-1)}{(N-1)}}$}, which, after normalizing for a
total wealth $W_T$ reads
\begin{eqnarray}
  w_R(t) =  W_T \ \frac{1-e^{-t\theta/N}}{1-e^{-t\theta}} e^{ - t  \theta \ \frac{(R-1)}{(N-1)}}.
\label{eq:10}
\end{eqnarray}
This expression is valid at long times, when ranks no longer change as
a result of economic exchange. At any fixed time, the ranked-wealth
distribution is thus exponential in rank. Accordingly, the transient
wealth distribution is of the form $P(w) \sim 1/w$. \Fig{fig:ginill}a
compares (\ref{eq:10}) with numerical results.
\subsubsection{Condensation criteria}
\label{sec:cond-crit}
In numerical simulations, a practical criterion is necessary to define
wealth condensation within predefined limits. We use for this purpose
the ratio $r(t)=w({2^{nd}})/w(1^{st})$, involving the wealths of the
richest and second-richest agents in the system. This ratio is one for
evenly distributed wealth, and goes to zero when all wealth condenses
onto a single agent. An alternative useful measure of condensation is
the normalized second moment
\begin{equation}
  \label{eq:4}
   W_2(t) =      \frac{ \sum _{i=1}^{N} w_i(t)^2}{\left (\sum _{i=1}^{N} w_i(t)\right )^2},
\end{equation}
which is similar to the participation ratio in localization
studies. $W_2$ is of order $1/M$ if wealth is more or less evenly
distributed among $M$ agents, and goes to one upon condensation onto a
single agent. Therefore, $1/W_2$ approximates the number of
economically active agents in the system, as much as the inverse
participation ratio estimates the number of sites over which a normal
mode, or an electron, spreads.
\subsubsection{Condensation timescales}
\label{sec:cond-timesc-cond}
The timescale $t_0(p,f,N)$ for convergence towards the condensed state
is an interesting property that quantifies the dynamics in the
unstable phase.  This timescale can be estimated theoretically, in the
full-mixture case. Using (\ref{eq:10}) one has that \hbox{$r(t) = e^{
    -t/t_0 }$}, with
\begin{equation}
  \label{eq:7}  
  t_0(p,f,N) \approx \frac{ N}{\theta(p,f)}.  
\end{equation}
From (\ref{eq:2}) and (\ref{eq:17}), we see that \hbox{$\theta \sim
  p^*(f)-p$}. Therefore, the condensation timescale $t_0$ diverges as
$(p^*(f)-p)^{-1}$ on approach to the critical interface.
\\
Simple analysis of (\ref{eq:10}) shows that $w_R(t)$ attains its
maximum value at time $T_R=t_0(p,f,N) \log{(R/(R-1))}$, and goes
exponentially fast to zero afterwards for all $R>1$. This is
understood in the following terms. During the condensation process in
the unstable phase, an agent with rank $R$ systematically extracts
wealth from poorer agents (those with $R'>R$) and transfers some of it
to richer agents (those with $R'<R$). As long as the wealth of poorer
agents so allows, his balance will be positive, so his wealth will at
first increase. But this increase happens at the expense of poorer
agents, and for times $t \approx T_R$ these will have exhausted their
wealth. Continued transference of wealth upwards (to richer agents)
will in turn make the agent with rank $R$ impoverish as well.
Therefore, each rank goes bankrupted at a specific timescale.  Poorer
ranks (of order $N$) do so at times of order $1/\theta$, while richer
ones (of order one) take time $t_0=N/\theta$. The second-richest agent
goes bankrupted at time $T_2= t_0 \log 2$, leaving a single rich agent
to account for most of the wealth. This justifies our identifying of
$t_0$ as the timescale needed for complete condensation. The entire
process of enrichment followed by bankruptcy for the different ranks
is visualized in \Fig{fig:ginill}.
\section{Network YS in the stable phase}
\label{sec:numer-results-stable}
\begin{figure}[htb]
\includegraphics[width=0.70\linewidth,angle=270]{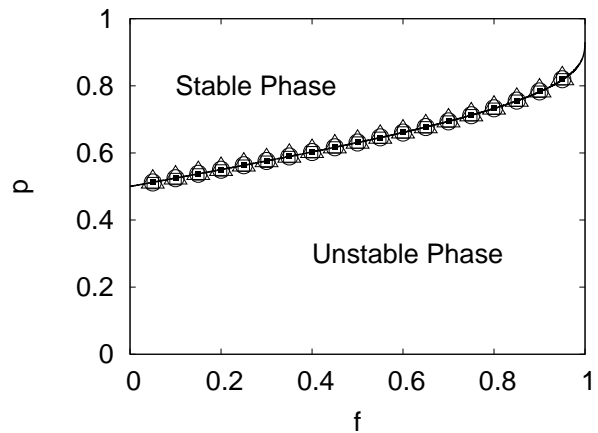}
\caption{ Critical probability $p_c(f)$ as obtained from fitting
  (\ref{eq:tau0pf}) to data for $\tau_0(p,f)$ in the full-mixture case
  (full squares), one-dimensional network (empty squares),
  two-dimensional networks (circles), and random graphs with
  $\gamma=10$ (triangles). Error bars are smaller than the symbols and
  were not drawn. The full line is the full-mixture theoretical
  prediction $p^{*}(f)$ as given by (\ref{eq:17}). The number of
  agents was $N=400$ in all cases. }
 \label{fig:interface}
\end{figure}

In this section, results from numerical simulations for 1d rings, 2d
square lattices with periodic boundaries, Erd\H{o}s-R\'enyi Random
Graphs, and full-mixture, are described and compared with analytic
predictions for the full-mixture case.  Starting from an even
distribution of wealth among the $N$ agents, the system is first
equilibrated during $T_{eq}$ timesteps before measurements are
taken. The required number of equilibration steps is determined by
measuring $c(\tau)$ for a series of increasing $T_{eq}$ values, until it
is found to no longer depend on $T_{eq}$.  System sizes from $N=100$
to $1000$ agents are considered.
\\
We first describe how the critical line $p_c(f)$ is determined
numerically in this work.  Firstly, after equilibrating the system as
described above, correlation functions $c(\tau)$ are measured in the
stable phase for many pairs $p,f$ and for each network considered.
Once $c(\tau)$ is known for each pair $(f,p)$ and for each network,
(\ref{eq:12}) is fitted to these data, from where estimates for the
relaxation times $\tau_0(p,f)$ are obtained.  As expected, $\tau_0$ is
found to diverge on approach to a critical value $p_c(f)$ that
delimits the stable phase from below.  By next fitting
(\ref{eq:tau0pf}) to our data for $\tau_0(p,f)$, we can obtain very
precise estimates for $p_c(f)$, the location of this divergence. Our
results are shown in \Fig{fig:interface}. The critical values so found
are, in all cases, consistent with the full-mixture prediction
(\Eqn{eq:17}) within numerical errors, suggesting that $p_c(f) =
p^{*}(f)$, for all networks considered.

The above result differs from expectations based on the theory of
equilibrium phase transitions. In that case, for a given interacting
system, critical parameters as e.g~the critical temperature, do depend
on the network, i.e.~are not universal. The analogous parameter for
Yard-Sale is the critical probability $p_c$, which, within our
numerical errors, seems to be network-independent and the same as in
the full-mixture, or Mean-Field, case. We therefore propose that the
critical interface (\ref{eq:17}) derived for the full-mixture case is
exact on any singly-connected network.
\\
Equilibrium wealth distributions in the stable
phase~\hbox{($p>p^{*}(f)$)} where measured (not shown) for all
networks considered in this work, for several pairs $(p,f)$, and
compared with full-mixture. We found that $P(w)$ is network-dependent,
although differences with full-mixture are minor. Relaxation
timescales, on the other hand, are found to be strongly
network-dependent, which is reasonable since the paths through which
wealth can flow are dictated by network topology. As expected,
relaxation to equilibrium takes longer on 1d rings, because there are
lesser paths for wealth to flow, and it is fastest in the full-mixture
case.
\section{Network YS in the unstable phase}
\label{sec:numer-results-unst}
\begin{figure}[htb]
\includegraphics[width=0.70\linewidth,angle=270]{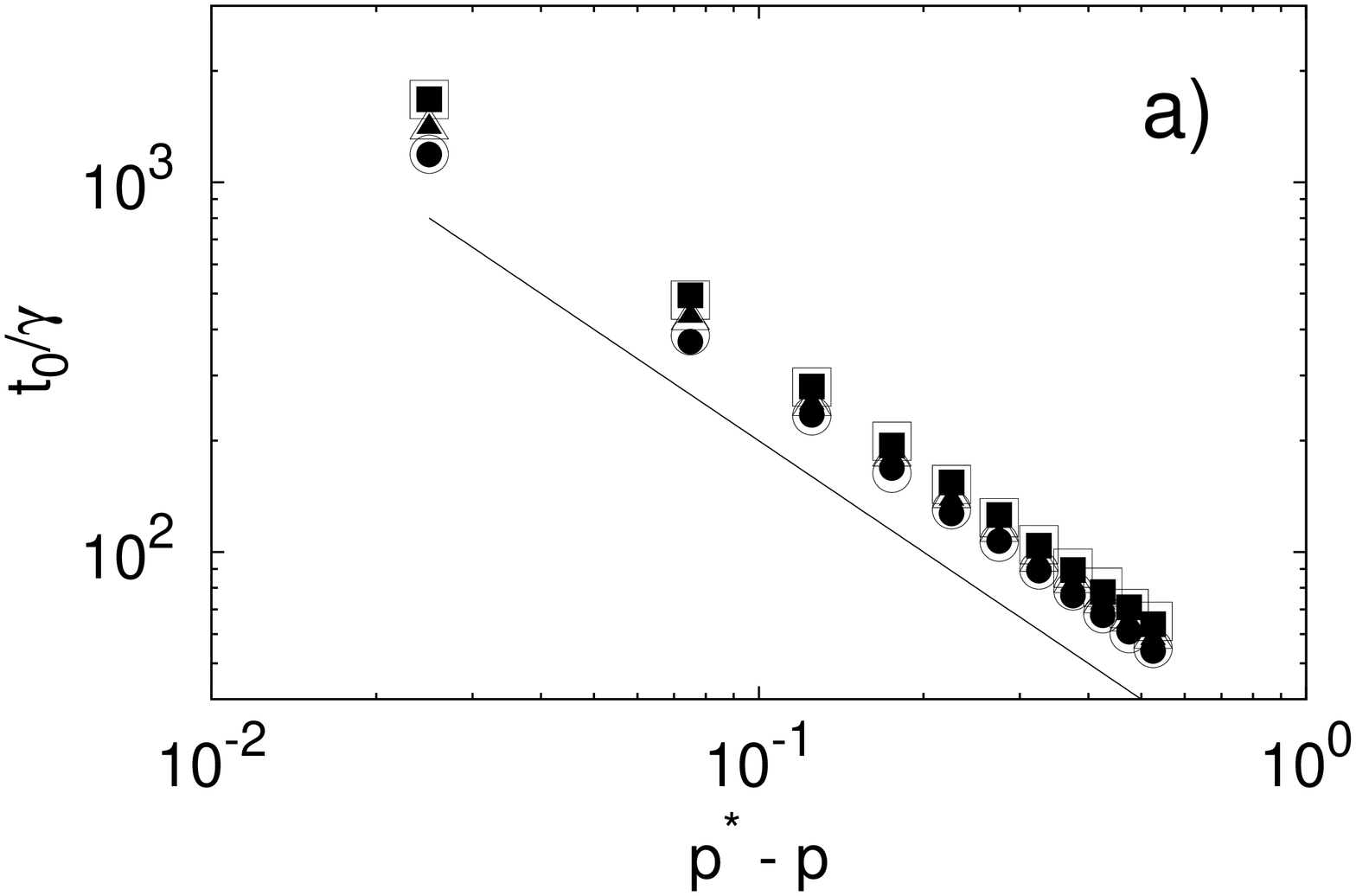}
\includegraphics[width=0.70\linewidth,angle=270]{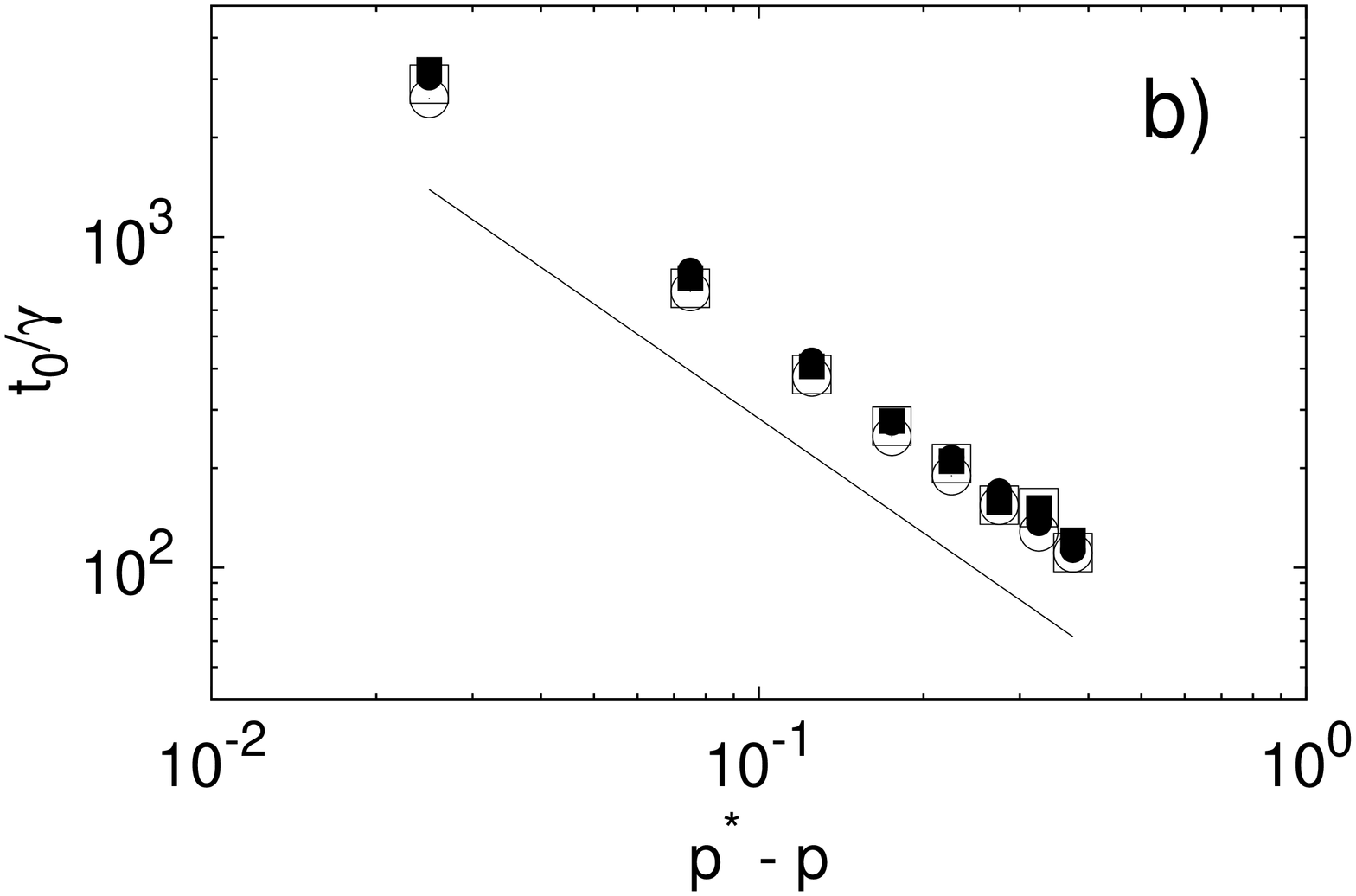}
\caption{ a) Condensation time $t_0$ divided by coordination $\gamma$,
  for random graphs with link density $\lden=\gamma/(N-1) =0.2$
  (squares) $0.8$ (triangles), and $1.0$ (full-mixture case,
  circles). Empty symbols show data for $N=400$, filled ones for
  $N=900$ agents.  The solid line is $ \sim 1/(p^*-p)$.  b)
  Condensation time $t_0$ divided by coordination number $\gamma =
  2d$, for periodic rings (squares) and periodic square lattices
  (circles) with $N=400$ (empty symbols) and $N=900$ (filled symbols).
  The exponent for the divergence at $p^*$ is estimated as $1.15 \pm
  0.2$ (solid line is \hbox{$ \sim 1/(p^*-p)^{1.15}$}.). }
 \label{fig:tcondensation}
\end{figure}
In the fully mixed case, whenever the system is in the unstable phase
$p<p^*(f)$, where $p^*(f)$ is given by (\ref{eq:17}), all wealth ends
up being owned by a single rich agent in the long run. This process is
called wealth condensation~\cite{MMAE11,MGIWCI07}.  For numerical
purposes, in this work we assume that the system is completely
condensed when $r(t)=w({2^{nd}})/w(1^{st}) \leq 10^{-4}$.  The time
$t_0$ needed for this limit to be reached is measured and averaged
over $10^3$ condensation histories. Results for $t_0/N$ in the
complete-graph limit (full mixture) are displayed in
\Fig{fig:tcondensation}a (open and filled circles), and are found to
behave as \hbox{$t_0(p,f) \propto (p^{*}(f)-p)^{-1}$}, in entire
accordance with the theoretical result (\ref{eq:7}) for full-mixture.

For network-restricted Yard-Sale in the unstable phase, complete
wealth condensation onto a single agent is no longer
observed. Instead, in the long run, the whole wealth condenses onto an
extensive set of locally rich agents (LRA). A locally rich agent is
\emph{defined} to be one who is richer than any of its
neighbors. Agents who are non-LRA impoverish steadily in the long run,
because in the unstable phase there is a systematic transference of
wealth from poor to rich agents.  For long times, each LRA is only
connected to agents whose wealth is extremely small. Wealth exchange
is then effectively suppressed, leading to dynamical freezing, onto a
disordered final state.
\begin{figure}[h!]
\begin{center}
 \includegraphics[width=0.70\linewidth,angle=270]{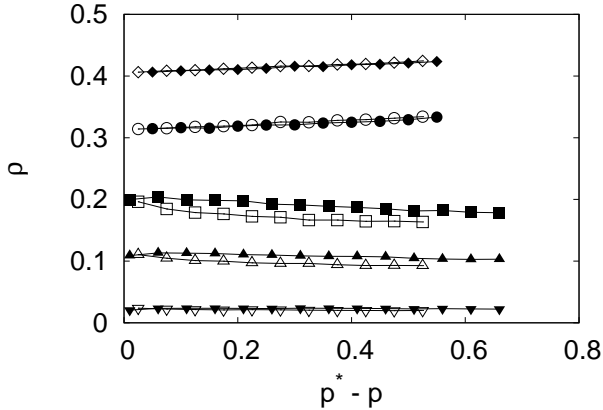}
 \caption{Final density $\rho$ of locally rich agents, as a function
   of distance $(p^*-p)$ to the interface, on 1d rings (diamonds), 2d
   square lattices (circles), and Random Graphs with average
   coordination $\gamma=10$ (squares), $20$ (triangles), and $100$
   (inverted triangles).  Empty symbols indicate data for for $f=0.1$
   and filled symbols for $f=0.2$. The number of agents was $N=400$ in
   all cases. }
 \label{fig:rholra}
\end{center}
\end{figure}
\subsection{Condensation times}
\label{sec:cond-times-locally}
At long times, there is a clear scale separation between the wealth of
each LRAs and those of its neighbors, the latter going exponentially
to zero in time.  We assume that the final set of LRAs has been
irreversibly frozen, and that wealth exchange has effectively stopped,
when each LRA is richer than its richest neighbor by a factor of at
least $10^{-4}$.  This criterion generalizes the one we adopted for
full-mixture, and reduces to it whenever there is condensation, in
which case there is only one LRA.

\Fig{fig:tcondensation}a shows condensation times divided by
coordination $\gamma$, for Random Graphs with variable link density
\hbox{$\lden = \gamma / (N-1)$}. The case $\lden=1$ (circles) is the
complete graph, or full-mixture case. Condensation times are seen to
diverge at the interface as $1/(p^*-p)$, that is, the exponent of this
divergence does not depend on $\gamma$ and is the same as for
full-mixture.  Our results also show that $t_0$ is roughly
proportional to $\gamma$, which is consistent with \Eqn{eq:7} for the
full-mixture limit, in which case $\gamma = N-1$.

\Fig{fig:tcondensation}b shows $t_0/\gamma$ for 1d and 2d networks,
where \hbox{$\gamma=2d$}. The exponent $\zed$ in \hbox{$t_0/\gamma
  \propto (p^{*}(f)-p)^{-\zed}$} seems to be slightly larger for these
finite-dimensional networks. Although $\zed$ is arguably
dimension-dependent, the quality of our data does not allow us to
resolve the difference between $\zed_{1d}$ and $\zed_{2d}$. Our best
estimate is $\zed_{1d,2d} = 1.15 \pm 0.20$ in one and two
dimensions.
\subsection{Locally Rich Agents}
\label{sec:locally-rich-agents}
Clearly, any set of LRAs with arbitrary wealths, surrounded by
impoverished agents is a fixed point of the dynamics.  There is thus a
non-denumerable multiplicity of fixed points, among which the wealth
exchange dynamics chooses one stochastically.  The statistical
properties of these fixed points, as for example the average number of
LRAs, and their wealth distribution, depend on the parameters of the
model, as well as on the topology of the network, among other
things. A detailed study of these properties is beyond the scope of
this work. However, some of the most relevant properties of LRAs,
namely their number and wealth distribution, will be briefly discussed
in the following.
\subsubsection{Number of LRAs}
\label{sec:number-lras}
Once the above described criterion for the formation of a set of LRAs
is satisfied, the dynamics is stopped, and the properties of LRAs are
determined. Measurements are averaged over $10^3$ repetitions of the
condensation history, for each case.

Condensation of wealth onto a reduced set of agents is a consequence
of the unstable nature of the dynamics for $p<p^*(f)$. There is a
systematic transfer of wealth from poor to rich, which in turn
increases wealth differences. The strength of this instability is
given by $\theta(p,f)$ (\Eqn{eq:2}), and becomes zero right at the
interface $p=p^*(f)$. Close to this interface, where $\theta$ is
small, wealth appropriation by the richer agents happens very slowly.
Wealth has then more time to migrate to richer agents, before the
dynamics comes to a halt. One therefore expects the process of wealth
concentration onto a single rich agent to happen more completely
there, than deep inside the unstable phase, where dynamical arrest
takes place in a short time. One could then expect the average number
of LRAs to decrease on approach to the interface.

However, our results, displayed in \Fig{fig:rholra}, show that, for a
given network, the final density $\rho$ of LRAs depends only very
mildly on the exchange parameters $p$ or $f$. In other words, the
number of LRAs is roughly the same in the whole unstable phase, and is
only determined by the network's properties (see
\Sec{sec:analyt-pred-rho}).  For 1d rings and 2d square lattices,
there is even no observable \hbox{$f$-dependence} in \Fig{fig:rholra}.
\subsubsection{Analytical prediction for $\rho$}
\label{sec:analyt-pred-rho}
\begin{figure}[h!]
\begin{center}
 \includegraphics[width=0.70\linewidth,angle=270]{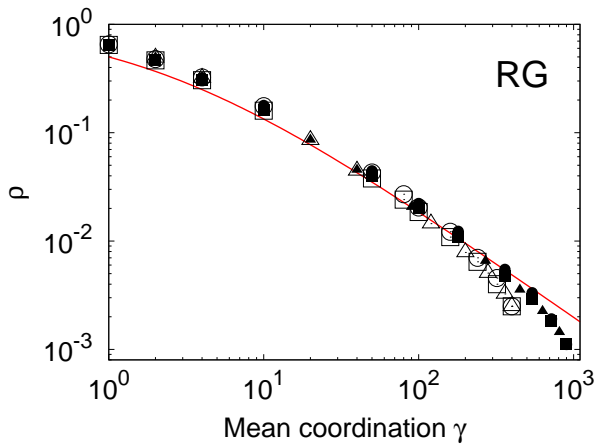}
 \caption{Final density $\rho$ of locally rich agents versus average
   coordination $\gamma$, on Random Graphs with $N=400$ agents (empty
   symbols), and $N=900$ agents (solid symbols), for $f=0.1, p=0$
   (squares), $f=0.6, p=0$ (circles), and $f=1.0, p=1/2$ (triangles).
   The solid line is a prediction from Abad~\cite{AO04}.}
 \label{fig:gamavsrichs}
\end{center}
\end{figure}

By definition, no two LRAs are connected to each other.  Therefore, a
set of LRAs constitutes an independent set~\cite{BRG01} of the graph.
A partition of a graph into independent sets constitutes a
coloring. We thus conclude that long-term YS evolution in the
appropriation phase identifies colorings of the network.

Since, as our numerical results suggest (see \Fig{fig:rholra}), the
number of LRAs is not strongly dependent on $f$ and $p$, one can
obtain useful information by studying the particular case $f=1$,
$p=1/2$, which is analytically tractable to some extent.  In this
particular case, whenever two agents interact, the winner is chosen at
random. If the richest agent wins, he gets the whole wealth of the
loosing agent, who is in turn rendered inactive. A similar process is
studied in the context of ``coagulation'' or ``coalescence'' \hbox{$A
  + A \to A + S$} of immobile reactants on a
network~\cite{BVNR99,AGNN01,AO04}.  Analytical descriptions for the
density of the active species $A$ ($S$ is the inert species), which in
our case is the final density of LRA, have been provided for these, as
well as for related models~\cite{KVA81,MPA93} that consider
``annihilation'' $A + A \to S+S$ as well.
\\
In particular, Abad~\cite{AO04} provides explicit expressions for the
final density $\rho$ of active agents on 1d and 2d lattices, as well
as on Bethe Lattices with coordination $\gamma$. If $\rho_0$ is the
initial density of active sites, the final density on a Bethe lattice
is
\begin{equation}
  \label{eq:3}
  \rho = \rho_0 \left ( 1 + \frac{\gamma -2}{2} \rho_0 \right )^{-\gamma/(\gamma-2)}.
\end{equation}
For large $\gamma$, this gives $\rho = \rho_0$ for \hbox{$\rho_0 <
  2/\gamma$}, and $\rho \sim 2/\gamma$ if \hbox{$\rho_0 > 2/\gamma$} .
\\
A comparison between (\ref{eq:3}) and our own numerical results on
Random Graphs (RG) with average coordination $\gamma$ is shown in
\Fig{fig:gamavsrichs}.  Notice that all sites on a Bethe lattice have
$\gamma$ neighbors, while this is satisfied only on average for Random Graphs.
Therefore, a perfect coincidence is not expected.  Nevertheless, an
acceptable similarity between our numerical results and (\ref{eq:3})
is found.

The 1d case is obtained from (\ref{eq:3}) in the limit $\gamma \to
2^+$, and equals $\rho=1/e$ for $\rho_0=1$ as is our case. Our
numerical result in 1d is approximately $0.4$ ( \Fig{fig:rholra}),
somewhat larger than this analytic prediction. Abad's two-dimensional
approximate result is $\rho=1/4$ for $\rho_0=1$, again slightly
smaller than our numerical results on 2d square lattices, shown in
\Fig{fig:rholra}.

These results show that the dynamical process of multiplicative wealth
concentration on networks has features in common with annihilation and
coalescence~\cite{KVA81,MPA93,BVNR99,AGNN01,AO04} of immobile
reactants.  Furthermore, it was in the context of those models that
the failure of the MF approximation to predict the final density of
active species was first noticed. While MF predicts a zero asymptotic
density of the active species, on generic networks a finite value is
found. This parallels our observation that, while in full mixture
wealth condenses onto a single agent, on networks it does so onto an
extensive set of agents.
\subsubsection{Wealth distribution of LRAs}
\label{sec:wealth-distr-lras}
\begin{figure}[h!]
\centerline{
 \includegraphics[width=0.35\linewidth,angle=270]{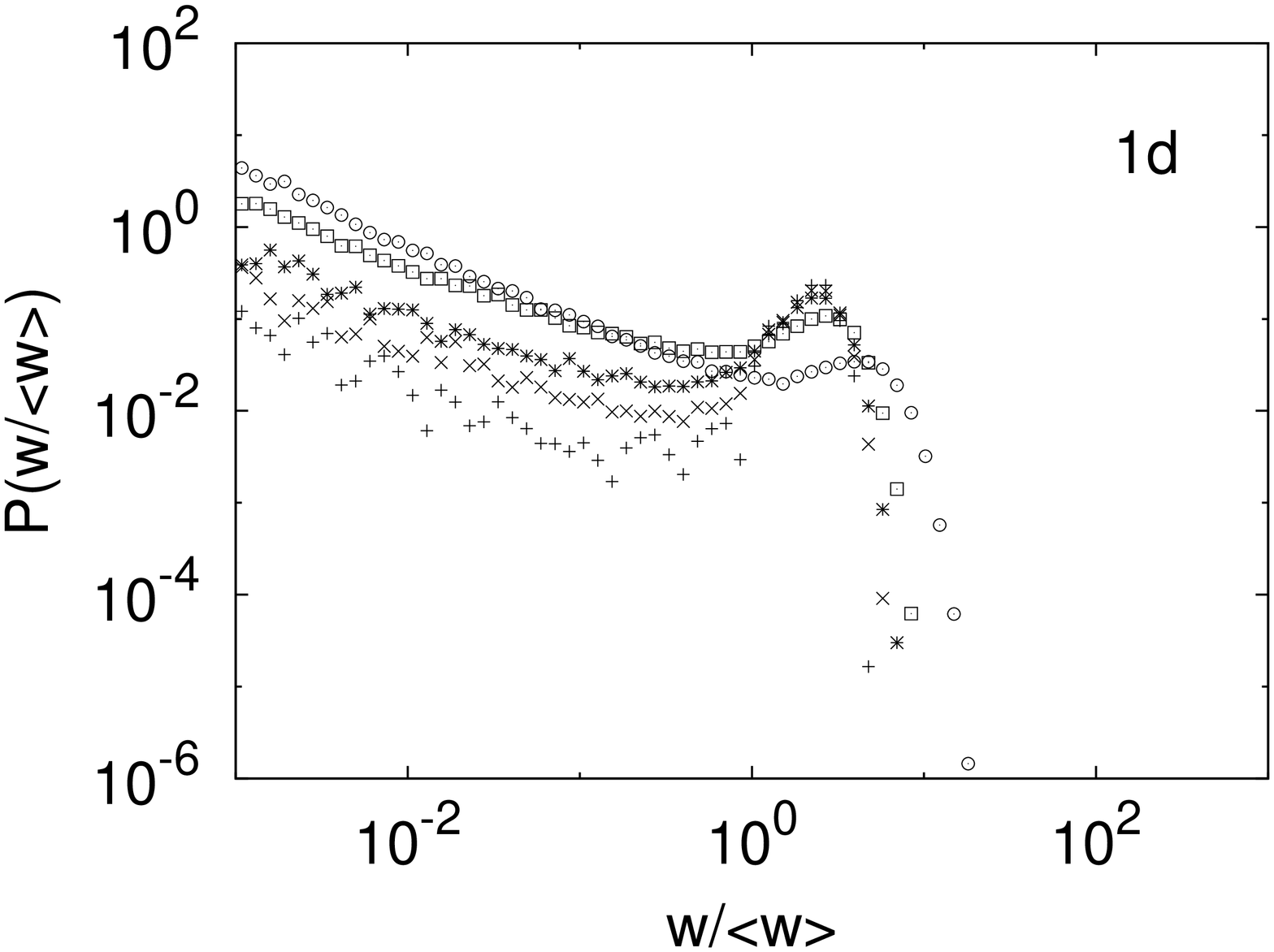}
 \includegraphics[width=0.35\linewidth,angle=270]{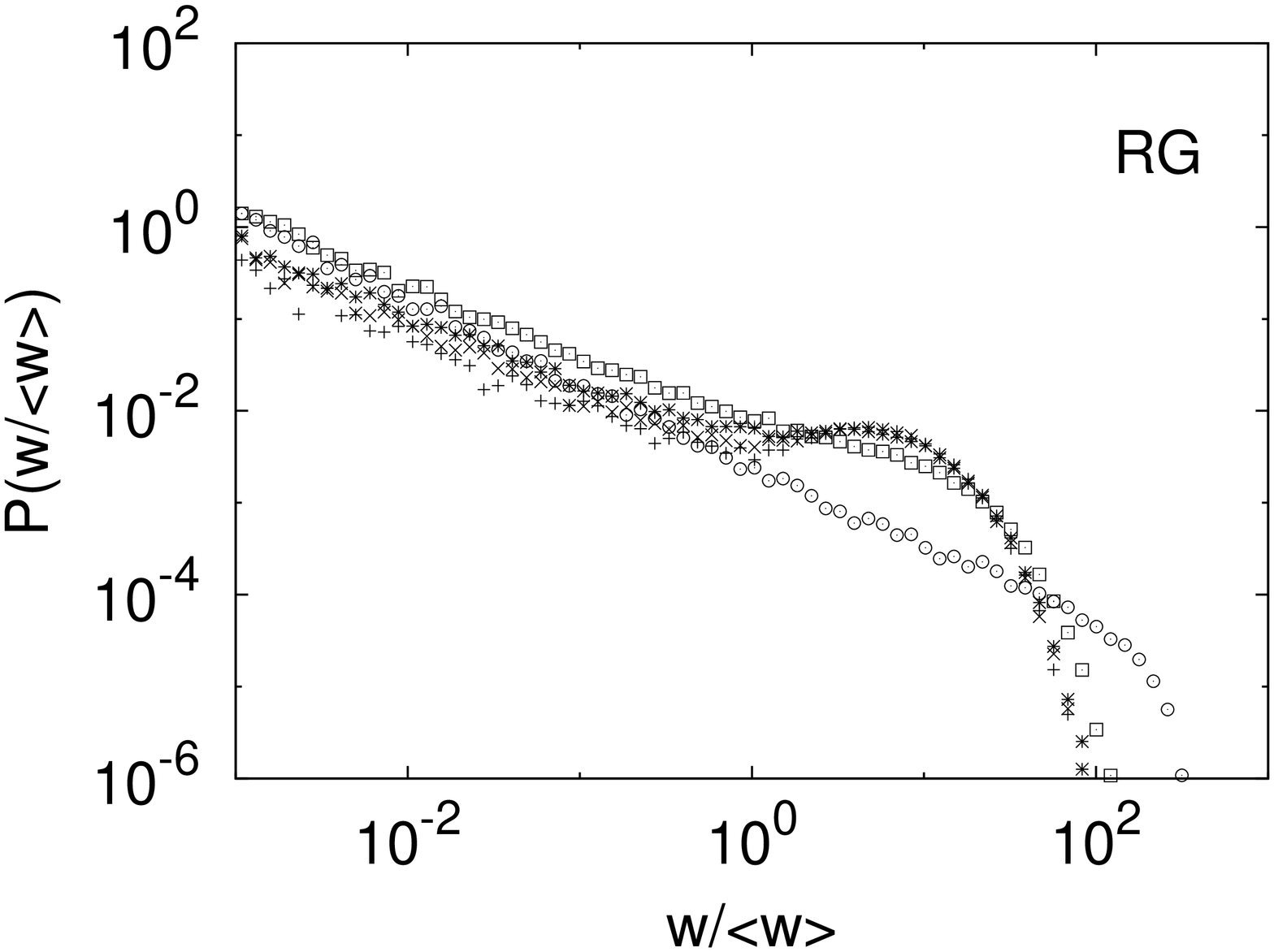}
}
\centerline{
 \includegraphics[width=0.35\linewidth,angle=270]{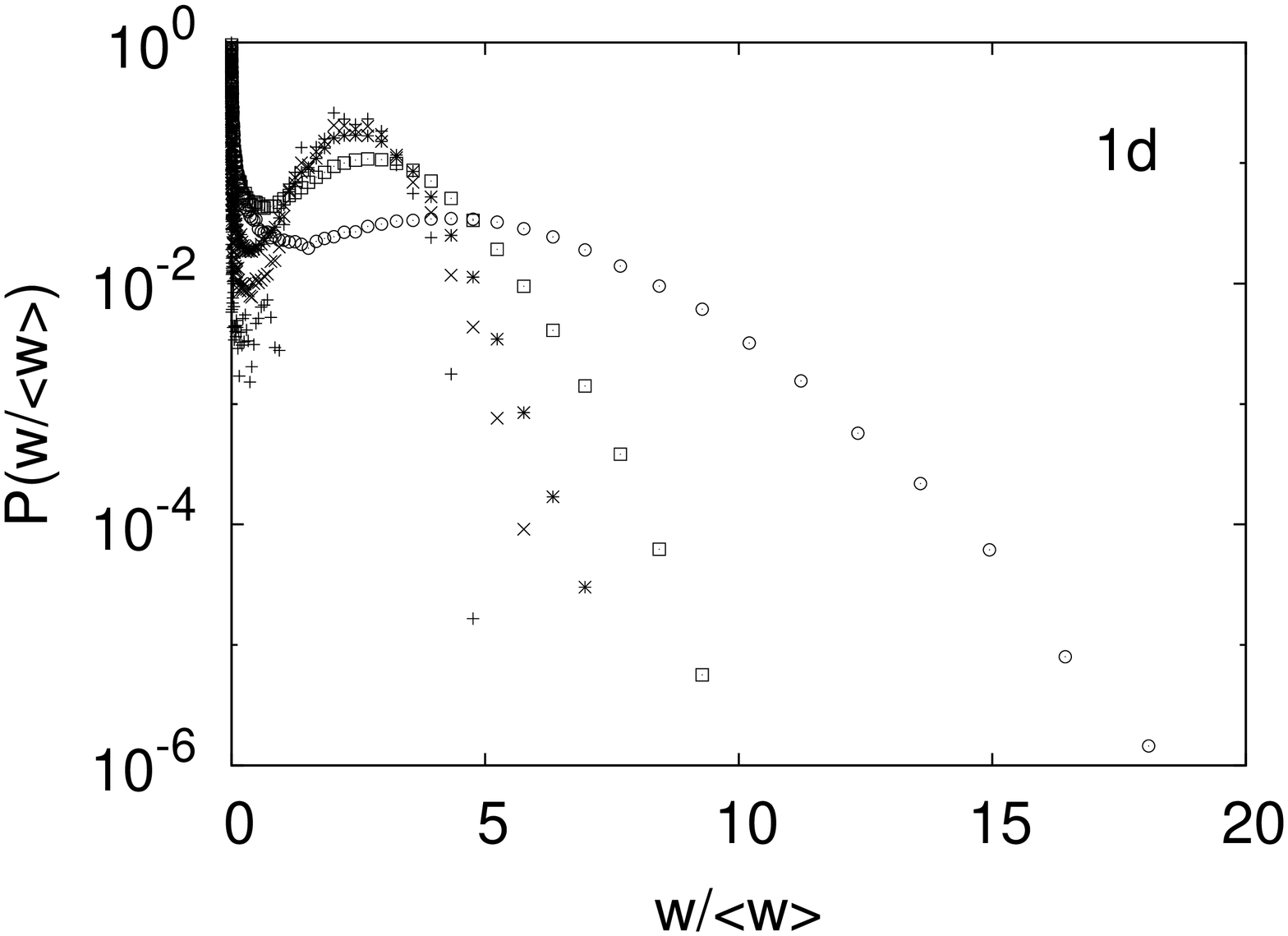}
 \includegraphics[width=0.35\linewidth,angle=270]{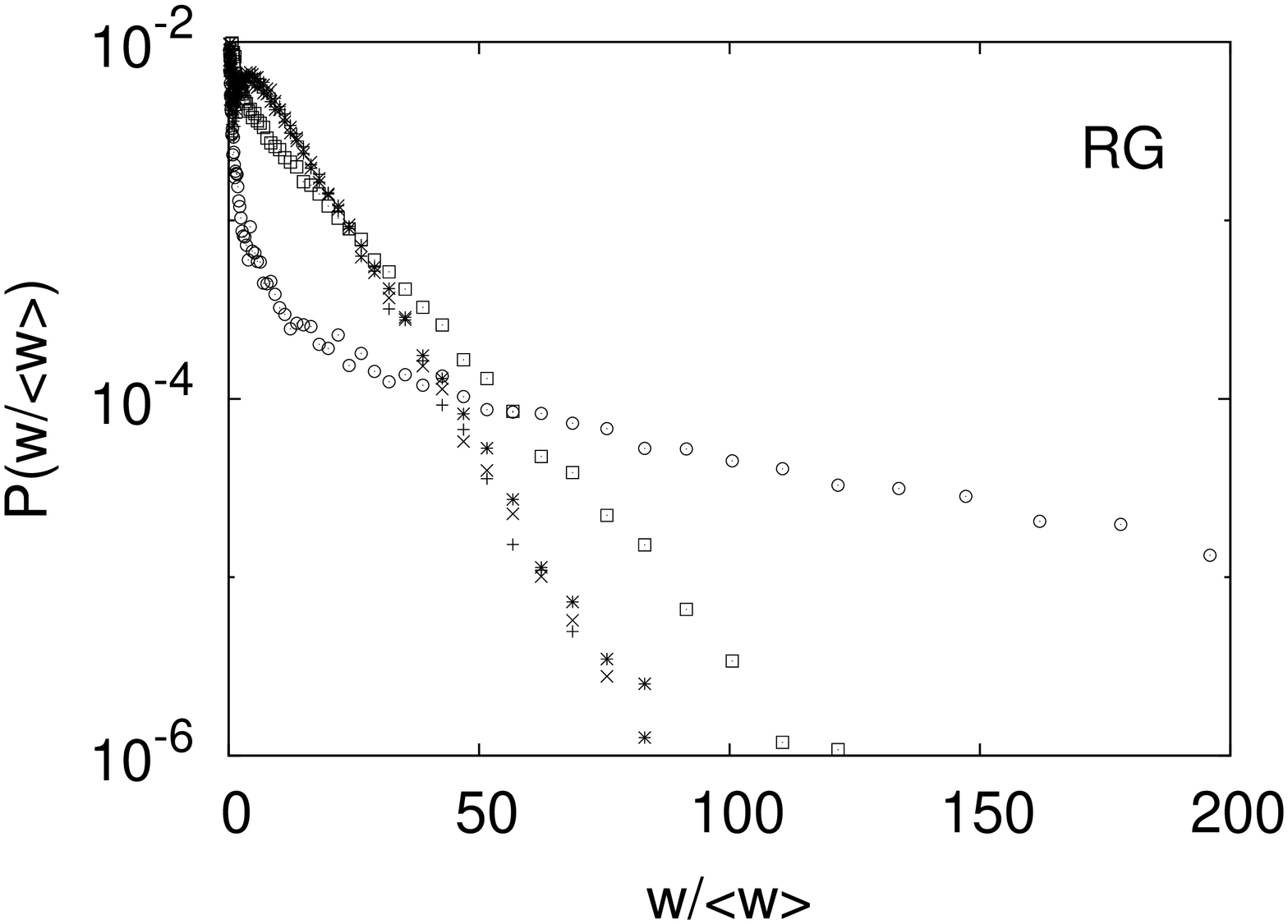}
}
\centerline{
 \includegraphics[width=0.35\linewidth,angle=270]{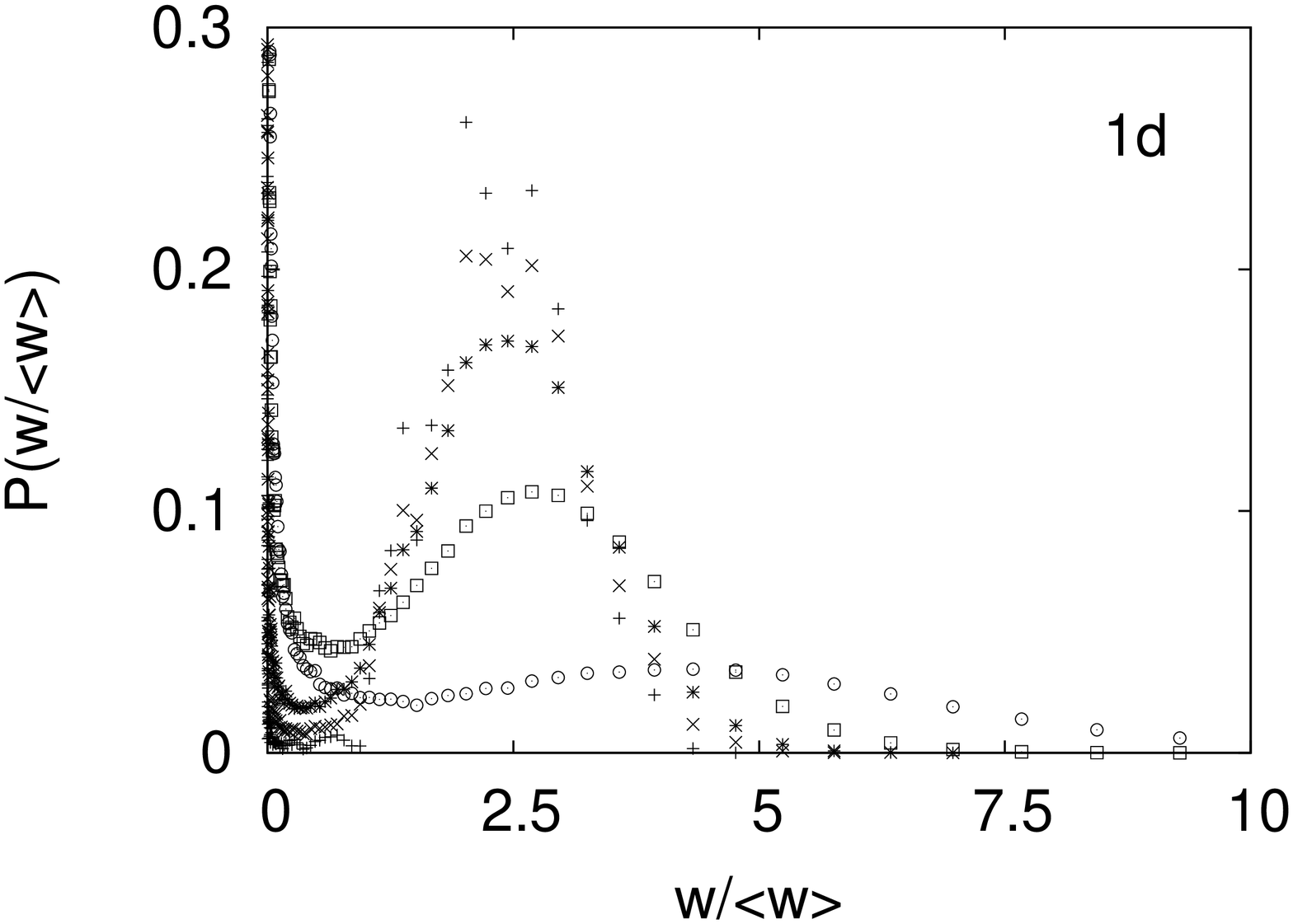}
 \includegraphics[width=0.35\linewidth,angle=270]{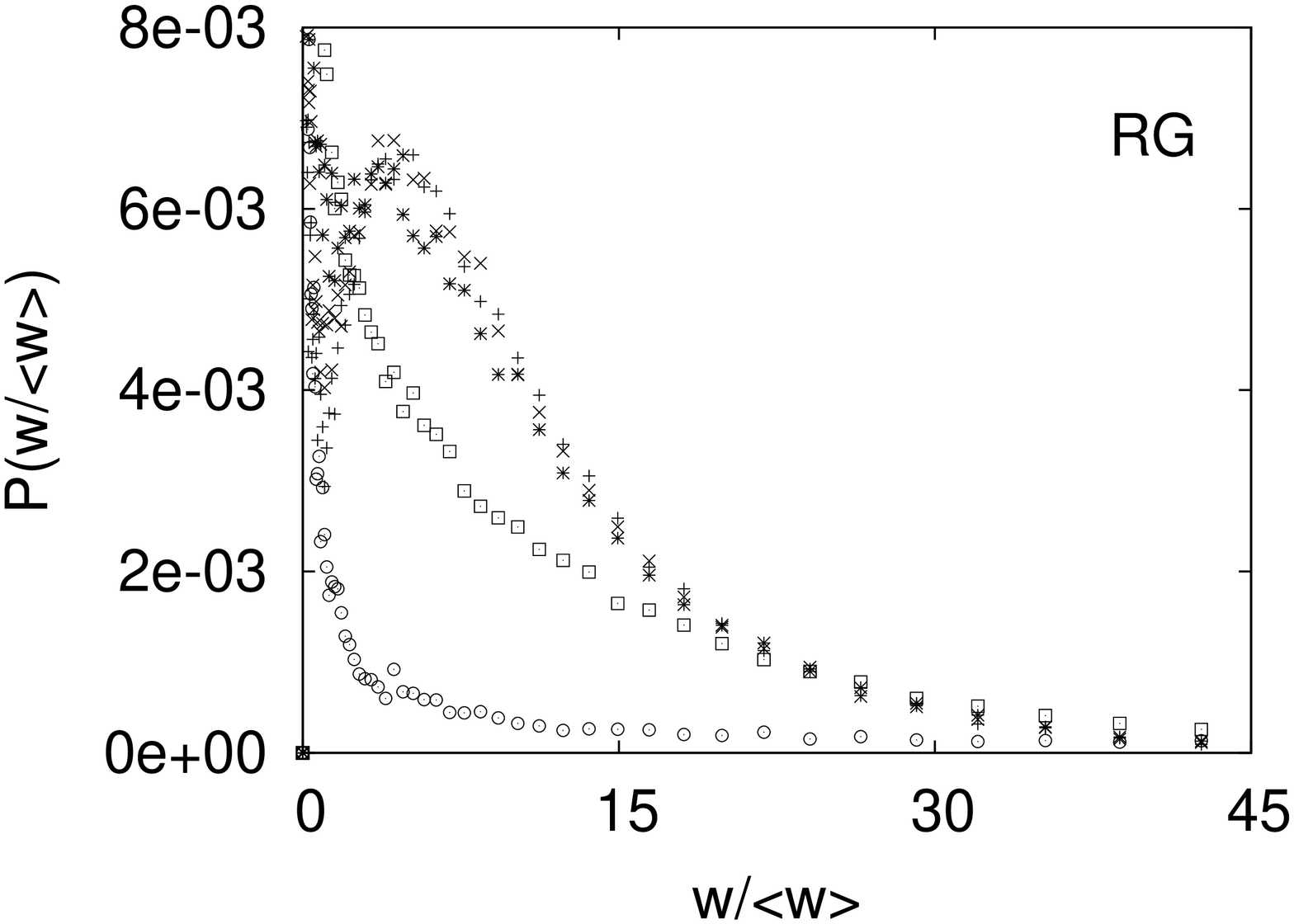}
}
\caption{Log-log (top row), log-lin (middle), and lin-lin (bottom)
  plots of wealth distribution of locally rich agents (LRA) in the
  frozen state, with $f=0.6$ and $p=$ 0 (plusses), 0.2 (crosses), 0.4
  (asterisks), 0.6 (squares), and 0.659 (circles). The critical
  probability for this value of $f$ is $p^*=0.660964$ in MF. The first
  column is from simulations on 1d rings (similar results were
  obtained on 2d square lattices) and the second one on Random Graphs
  with $\gamma=20$.  In all cases, $N=400$ agents was used.}
 \label{fig:richsdistr1d}
\end{figure}

\begin{figure}[htb]
\centerline{
  \includegraphics[width=0.70\linewidth,angle=270]{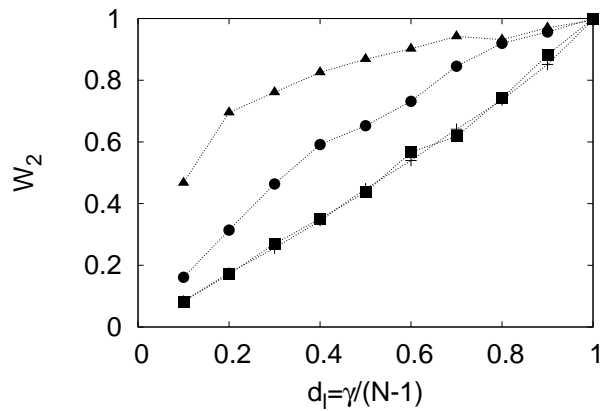}
}
\caption{ Second moment $W_2$ (\Eqn{eq:4}) of wealth distribution on
  Random Graphs at freezing, versus link density $\lden =
  \gamma/(N-1)$, for $f=0.1$ and $p=0.0$ (plusses), $0.45$ (squares),
  $0.52$ (circles), and $0.525$ (triangles). Dotted lines are guides
  to the eye. The critical interface for this value of $f$ is located
  at $p^*=0.525042$. The number of agents is $N=200$.  }
 \label{fig:w2.freezing}
\end{figure}

Wealth distributions of LRAs in the frozen state are displayed in
\Fig{fig:richsdistr1d} for 1d and RG with $\gamma=20$. Wealth
distributions on 2d lattices were also measured (are not shown) and
found to be qualitatively similar to those in 1d.  In all three cases,
the minimum in $P(w/\aver{w})$ for $w/\aver{w} \approx 1$ suggests the
existence of two sets of LRAs with different properties.  For the sake
of this analysis, the population of LRAs is divided in two groups
according to their wealth. Those with wealth \hbox{$w > \aver{w}$},
have a roughly normal wealth distribution in 1d and 2d, and an
exponential wealth distribution on Random Graphs. We call these ``type
1'' LRAs. In addition to those, LRAs with \hbox{$w < \aver{w}$}, have
wealths distributed according to a power-law $P(w) \sim 1/w$ that
extends down to zero. These we call ``type 2'' LRAs.
\\
We have measured (not shown) the numbers of type 1 and 2 LRAs versus
time for all networks with various parameter values. For $p$ values
that are not too close to the interface, freezing occurs rapidly, and
the final number and cumulative total wealth of type 2 LRAs turns out
to be almost negligible compared to those of type 1. In other words,
most LRAs are type 1, i.e.~have wealths larger than average in the
frozen state. Additionally, the wealths of type 1 LRAs are found to
have a narrow distribution if not too close to the interface. Very
close to the interface, i.e.~ for $p \to p^*(f)$, on the other hand, a
significant amount of conversion from type 1 to type 2 occurs before
the frozen state is reached. During this process, the number of type 1
LRAs drops steadily, while the total number of LRAs stays almost
constant or increases slowly. Conversion from type 1 to type 2 means
that a large number of LRAs, despite being richer than their
neighbors, can still loose a significant fraction of their wealth,
which in the end goes to the few remaining type 1 LRAs. This is
possible because close to the interface $\theta$ (\Eqn{eq:2}) is
small, and therefore being richer does not ensure a strong statistical
advantage.
\\
On approach to the interface, wealth distributions of LRAs develop
long tails for large wealth, and the power-law behavior $P(w) \sim
1/w$ is seen to extend to the right. Therefore, the distinction
between the wealth distributions of type 1 and type 2 LRAs is blurred
in this limit.  Near the interface, a single LRA ends up owning a
significant fraction of the whole wealth. Therefore, even though the
total number of LRAs remains approximately constant when the interface
is approached, most of them will only have negligible wealth in the
end. Therefore, we conclude that wealth condenses onto a single rich
agent, on any connected network, when the system is unstable but very
close to the critical interface $p^{*}(f)$.
\\
In the case of Random Graphs, condensation onto a single agent occurs
in the whole unstable phase only in the complete graph limit, i.e.~in
the limit $\lden = \gamma/(N-1) \to 1$. A measure of wealth
condensation is provided by $W_2$ (see \Eqn{eq:4}). A plot of $W_2$ at
freezing versus link density $\lden$ is shown in
\Fig{fig:w2.freezing}. These data show that full condensation in the
whole unstable phase only happens in the complete-graph limit. Inside
the unstable phase, wealth is distributed among all LRAs roughly
uniformly. Unless the system is really close to the interface, one has
$W_2 \sim \lden$.  This result can be understood as follows. In the
frozen state, as shown in \Fig{fig:richsdistr1d}, wealth is
distributed exponentially among the resulting $N_{LRA}(\gamma)$
locally rich agents. If there are $N_{LRA}$ locally rich agents and
the rest have zero wealth, \Eqn{eq:4} can be rewritten as \hbox{$W_2
  =1/N_{LRA} (1+\sigma^2/\aver{w}^2)$}, where $\sigma$ is the variance
of the wealth distribution of the LRAs.  For the particular case of an
exponential distribution, \hbox{$\sigma^2=\aver{w}^2$} and therefore
$W_2 = 2/N_{LRA}$. As discussed in \Sec{sec:analyt-pred-rho}, the
density $\rho$ of LRA is approximately $2/\gamma$ for large
$\gamma$. Therefore $N_{LRA} = 2N/\gamma = 2/\lden$, which renders
$W_2 \sim \lden$ as observed numerically for $(p,f)$ points not too
close to the interface (\Fig{fig:w2.freezing}).

\section{Discussion of results}
\label{sec:discussion-results}
Yard-Sale
(YS)~\cite{IKRWDI98,HFTM02,SSMW03,IGACBR04,SPWATM04,MGIWCI07,MMAE11}
is a simple but realistic model for commercial exchange that presents
two phases: a stable (or wealth-sharing) phase where wealth is
distributed and an unstable (or wealth-appropriation) one where wealth
concentrates in the hands of a few agents. We have numerically studied
the static and dynamic properties of YS on several types of networks,
comparing them to those in the full-mixture (or mean-field)
approximation.  Equilibrium wealth distributions $P(w)$ on networks,
in the stable phase, are found to be very similar to those in
full-mixture. Measuring decorrelation times $\tau_0$, which in the
this model can be interpreted as ``social mobility'' times, we are
able to very precisely locate the interface that separates the wealth
sharing from the wealth appropriation phases
(\Fig{fig:interface}). Our numerical results strongly suggest that the
critical interface $p^{*}(f)$ derived in the full-mixture
approximation (\Eqn{eq:17}) is exact on any network.
\\
An important result is the observation that, for network YS in the
unstable phase, wealth does not condense onto a single agent as it
does in the fully mixed case, but onto an extensive set of locally
rich agents (LRA) instead. These LRAs form an independent
set~\cite{BRG01} in the network, and therefore define a coloring of
it. The final density of agents with nonzero wealth is thus finite on
networks, while it is zero for full mixture. In recent related
work~\cite{S-MHRCFI11}, it was proposed that the emergence of many
locally rich agents might be due to multiple-connectedness of the
network, suggesting that, on networks made of just one connected
component, global condensation onto one agent would eventually occur.
This expectation is not confirmed by our results, which show that an
extensive number of LRAs remain, in the whole unstable phase, on
singly-connected networks as well. It is only in the limit of a
complete graph, which is the fully mixed case, or, (on any network) in
the limit $\theta \to 0$ (i.e.~right at the interface), that
condensation onto a single agent is observed.
\\
We have discussed previously unnoticed connections between wealth
condensation in YS and earlier studies of
annihilation~\cite{KVA81,MPA93} or
coalescence~\cite{BVNR99,AGNN01,AO04} of immobile reactants, a related
statistical problem where the distinction between network results and
mean-field ones (i.e.~zero vs nonzero final density of LRAs) was first
noticed~\cite{BVNR99,AGNN01}. With the help of these connections, we
have been able to compare our own numerical results (\Figs{fig:rholra}
and \ref{fig:gamavsrichs}) with analytic predictions~\cite{AO04} for
the remaining density of wealth-possessing agents on several networks.
A good coincidence is found throughout the entire unstable
phase. Furthermore, by using analytical expressions for the density of
remaining LRAs on Random Graphs, we were able to explain our numerical
results (\Fig{fig:w2.freezing}) showing that \hbox{$W_2 \propto d_l$}
on Random Graphs, deep inside the unstable phase.
\\
Surprisingly, the density of LRAs is essentially constant in the whole
unstable phase, although their wealth distribution is not. Their
wealth distribution is roughly homogeneous, i.e.~has a relatively
narrow distribution, except when extremely close to the interface. A
particularity that deserves further attention is the fact that the
wealth distribution of LRAs in the frozen state is nearly normal in
one and two dimensions, but exponential on Random Graphs
(\Fig{fig:richsdistr1d})
\\
Very close to the interface that delimits the unstable phase from
above, however, wealth is no longer homogeneously distributed among
the LRAs, but develops a long right tail of the form $P(w) \sim 1/w$
until, at the interface itself, only one rich agent remains, which
owns the whole systems' wealth. Therefore, on the interface itself,
condensation onto a single agent is again observed, on any
network. However, the time needed for condensation diverges in this
limit, in contraposition to the full mixture case, where wealth
condenses onto a single agent in finite time.
\\
We have found that YS models only show strongly network-dependent
properties when the system's parameters $(p,f)$ are in the unstable
phase. In the light of this result, the correlations between
topological properties and wealth distributions that have been
recently observed in experimental studies of global commercial
networks~\cite{SBTOT03,GLFTP04,GD-MAIBT07}, may be interpreted as
suggesting that the international trade system is itself in the
unstable phase. In other words, that the microscopic exchange rules
for international trade are such that favor systematic wealth
appropriation by larger agents. Other evidences of this possibility
have been recently found by analyzing the distribution of per-capita
gross domestic products~\cite{Note4}.
\\
On the other hand, as already said in the introduction, global wealth
distributions depend on processes other than conservative
exchange. The generation of wealth by endogenous processes, for
example, acts as a source of wealth that would avoid freezing in the
unstable phase. A system under unstable exchange, in the presence of
wealth creation by endogenous processes, would reach a
quasi-stationary state in which wealth is produced everywhere and then
channeled towards richer agents by the exchange processes.
\\
\acknowledgements The authors wish to thank the warm hospitality of
the Statistical Mechanics group in Centro At\'omico Bariloche,
Argentina, where this work was started. Computer time provided by
CGSTIC-CINVESTAV on ``Xiuhcoatl'' hybrid supercomputing cluster made
this work possible. R.~B.~G.~Acknowledges financial support from
CONACYT.

\end{document}